\documentclass[%
 reprint,
superscriptaddress,
 amsmath,amssymb,
 aps,
]{revtex4-2}
\usepackage{graphicx}
\usepackage{dcolumn}
\usepackage{bm}
\usepackage{amsmath}
\usepackage{hyperref}
\usepackage{color}
\usepackage{multirow}
\usepackage{array}
\usepackage{booktabs}
\usepackage{mathptmx}%
\usepackage{upgreek}%
\usepackage{physics}
\usepackage{CJK}

\begin{document}
\begin{CJK*}{UTF8}{gbsn} 
\preprint{APS/123-QED}

\title{High-Performance Microwave Frequency Standard Based on Sympathetically
Cooled Ions}

\author{Hao-Ran Qin({\CJKfamily{gbsn}秦浩然})
}
\affiliation{
  State Key Laboratory of Precision Measurement Technology and Instruments, Tsinghua University, Beijing 100084, China
}%
\affiliation{
 Department of Physics, Tsinghua University, Beijing 100084, China
}

\author{Sheng-Nan Miao({\CJKfamily{gbsn}苗胜楠})
}%
\affiliation{
State Key Laboratory of Precision Measurement Technology and Instruments, Tsinghua University, Beijing 100084, China
}%
\affiliation{
  Department of Precision Instruments, Tsinghua University, Beijing 100084, China
}

\author{Ji-Ze Han({\CJKfamily{gbsn}韩济泽})
}
\affiliation{
State Key Laboratory of Precision Measurement Technology and Instruments, Tsinghua University, Beijing 100084, China
}%
\affiliation{
  Department of Precision Instruments, Tsinghua University, Beijing 100084, China
}

\author{Nong-Chao Xin({\CJKfamily{gbsn}辛弄潮})
}
\affiliation{
State Key Laboratory of Precision Measurement Technology and Instruments, Tsinghua University, Beijing 100084, China
}%
\affiliation{
  Department of Precision Instruments, Tsinghua University, Beijing 100084, China
}

\author{Yi-Ting Chen({\CJKfamily{gbsn}陈一婷})
}
\affiliation{
State Key Laboratory of Precision Measurement Technology and Instruments, Tsinghua University, Beijing 100084, China
}%
\affiliation{
  Department of Precision Instruments, Tsinghua University, Beijing 100084, China
}

\author{J. W. Zhang({\CJKfamily{gbsn}张建伟})
}
 \email{zhangjw@tsinghua.edu.cn}
 \affiliation{
  State Key Laboratory of Precision Measurement Technology and Instruments, Tsinghua University, Beijing 100084, China
  }%
  \affiliation{
    Department of Precision Instruments, Tsinghua University, Beijing 100084, China
  }

\author{L. J. Wang({\CJKfamily{gbsn}王力军})
}
 \email{lwan@tsinghua.edu.cn}
 \affiliation{
  State Key Laboratory of Precision Measurement Technology and Instruments, Tsinghua University, Beijing 100084, China
  }%
  \affiliation{
 Department of Physics, Tsinghua University, Beijing 100084, China
}
  \affiliation{
    Department of Precision Instruments, Tsinghua University, Beijing 100084, China
  }

\date{\today}

\begin{abstract}
  The ion microwave frequency standard is a candidate for the next generation of microwave frequency standard with the potential for very wide applications.
  The Dick effect and second-order Doppler frequency shift (SODFS) limit the performance of ion microwave frequency standards. The introduction
  of sympathetic cooling technology can suppress the Dick effect and SODFS and improve the stability and accuracy of the frequency standard.
  However, the sympathetically-cooled ion microwave frequency standard has seldom been studied before. This paper reports the first sympathetically-cooled
  ion microwave frequency standard in a Paul trap. Using laser-cooled ${}^{40}\mathrm{{Ca}}^{+}$ as coolant ions, ${}^{113}\mathrm{{Cd}}^{+}$ ion crystal is cooled to below 100 mK and
  has a coherence lifetime of over 40 s. The short-term frequency stability reached $3.48 \times 10^{-13}/\tau^{1/2}$, which is comparable to that of the mercury ion
  frequency standard. Its uncertainty is $1.5\times 10^{-14}$, which is better than that of directly laser-cooled cadmium ion frequency standard.
\end{abstract}

\maketitle
\end{CJK*}

\section{Introduction}

Faced with long-distance, short-duration, and high-speed situations, atomic clocks play an extremely important
role in basic science and practical applications. Because of their compact volume and mature technology, microwave
frequency standards are widely used in satellite navigation \cite{bandi2011high, mal2010space}, telecommunications \cite{ho1998new}, timekeeping \cite{diddams2004standards,Hg2008} and space exploration \cite{burt2021demonstration, liu2018orbit, prestage2007atomic}. Trapped-ion clocks constitute promising candidates for the next generation
microwave atomic clocks due to the advantages of compactness, high transportability, and high performance. To date, ion frequency standards based on ${}^{199}\mathrm{{Hg}}^{+}$ \cite{Hg1990, Hg1998, Hg2008, burt2021demonstration}, ${}^{113}\mathrm{{Cd}}^{+}$ \cite{Cd2013, Cd2015, Cd2021}, ${}^{171}\mathrm{{Yb}}^{+}$ \cite{Yb2014,Yb2019, xin2022} have brought about considerable progress.

Soon after cooling technology was proposed \cite{h1975cool, a1978trap} and realized for the first time \cite{chu1986ex}, it was quickly and widely
applied in atomic clock devices \cite{dai2021cold}. Fountain clocks based on ultracold cesium atoms play an important role
in establishing primary frequency standards of the SI (the International System of Units) second \cite{UTC2019}. Optical
clocks based on laser-cooled single ion or atoms in optical lattice are expected to feature in establishing the next-generation
definition of the second \cite{2019unit}.

Previously, the lack of a suitable electric dipole transition energy level or a laser capable of generating a suitable
wavelength had limited the use of some atomic particles that could not be directly laser-cooled. As the result,
sympathetic cooling was introduced into the field of atomic clocks \cite{brewer2019al+, AlCa2019}.
With relative uncertainties below $10^{-18}$, single ion of $\mathrm{{}^{27}Al^+}$ sympathetically cooled by  $\mathrm{{}^{25}Mg^+}$ underlie the most accurate optical atomic clocks so far built \cite{brewer2019al+}.

Nevertheless, frequency standard based on laser-cooled trapped ions usually suffer from the Dick effect because
of the dead time in the laser-cooling process and second-order Doppler frequency shift (SODFS) introduced by
the ions rising temperature during interrogation. These limitations can be overcome by sympathetic cooling.
However, prior to our work, the only sympathetically-cooled ion microwave frequency standard realized was
that constructed by J. J. Bollinger and colleagues using a Penning trap with 
   $\mathrm{Be^+}$ sympathetically cooled by $\mathrm{Mg^+}$ \cite{Be1991}. In past
   work, we have demonstrated the laser-cooled cadmium ion microwave frequency standard \cite{Cd2013,Cd2015,Cd2021}, and have
   made much progress in sympathetically-cooled large ion crystals \cite{zuo2019, han2021, miao2022}.

In this work, we report for the first time the realization of a high-performance microwave frequency standard
based on sympathetically-cooled cadmium ions. Approximately five thousand ${}^{113}\mathrm{{Cd}}^{+}$ ions are trapped in a linear
Paul trap and are sympathetically cooled to below 100 mK with ${}^{40}\mathrm{{Ca}}^{+}$, forming a large two-component ion
crystal. This frequency standard exhibits a short-term fractional frequency stability of $3.48 \times 10^{-13}/\tau^{1/2}$ (where $\tau$ is the averaging time), with a fractional accuracy of $1.5\times 10^{-14}$.
Compared with a hydrogen maser, the short-term
stability of the clock signal is indirectly measured to be $1.36\times 10^{-13}/\tau^{1/2}$ obtained because of the ultra-long
Ramsey free evolution time of 50 s, which, encouragingly, ensures future improvements in performance.

\section{Realization of Sympathetic Cooling}
\begin{figure}
  \includegraphics[width=\linewidth]{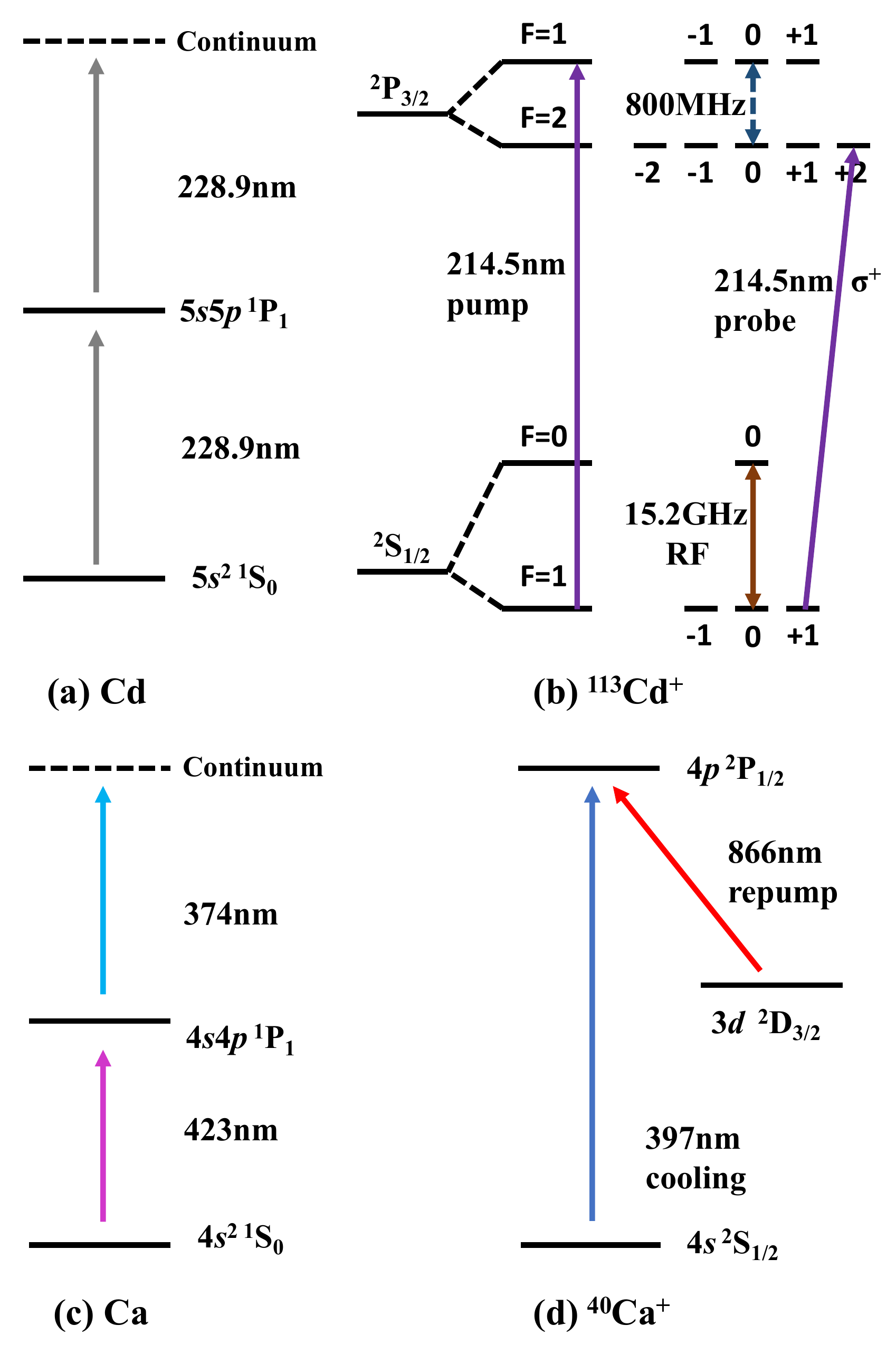}
  \caption{
    Relevant energy level structure of cadmium and calcium: (a) and (c) are the energy level structures of neutral atoms, while (b) and (d) are those of isotopic ions. The frequency or wavelength of the electromagnetic wave the associated with each transition  is given.
    }
  \label{fig:el}
\end{figure}

In our work, we choose laser-cooled (LC) ${}^{40}\mathrm{{Ca}}^{+}$ as coolant ions and sympathetically-cooled (SC) ${}^{113}\mathrm{{Cd}}^{+}$ as ``clock'' ions. The relevant energy level structure of cadmium and calcium are shown in Figure \ref{fig:el}. 
The ground state hyperfine splitting frequency (15.2 GHz) of the  ${}^{113}\mathrm{{Cd}}^{+}$ is the second largest among all working energy levels
of atomic microwave clocks, only to ${}^{199}\mathrm{{Hg}}^{+}$ with a frequency of 40.5 GHz. The 800 MHz ${}^{2}\mathrm{{P}}_{3/2}$ state hyperfine
splitting ensures that only one laser is needed to complete the state preparation and detection. Therefore, the ${}^{113}\mathrm{{Cd}}^{+}$ frequency standard has promise for high performance in application as well as potential for miniaturization
in fabrication. We adopted ${}^{40}\mathrm{{Ca}}^{+}$ as coolant ions because the natural abundance of ${}^{40}\mathrm{{Ca}}$ is as high as 97\%, and hence easily available. Moreover, the laser providing the cooling is also readily available. Compared with
previously used ${}^{24}\mathrm{{Mg}}^{+}$ \cite{zuo2019}, ${}^{40}\mathrm{{Ca}}^{+}$ has a closer mass to ${}^{113}\mathrm{{Cd}}^{+}$, which facilitates sympathetic cooling.

\begin{figure}
  \includegraphics[width=\linewidth]{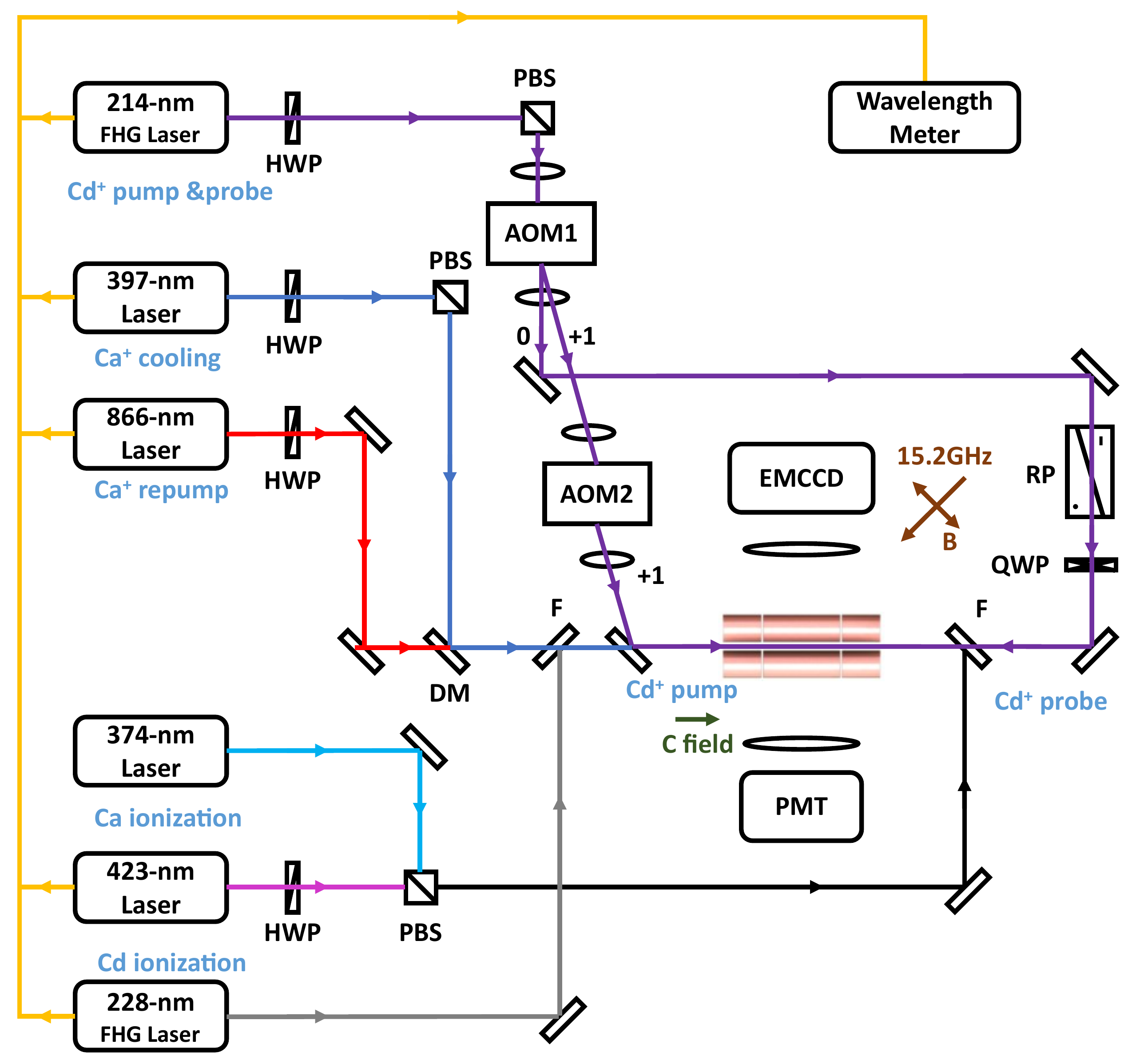}
  \caption{
    Schematic of the entire experimental system. HWP: half-wave plate; QWP: quarter-wave plate; PBS: polarization beam splitter; DM: dichromatic mirror; RP: Rochon polarizer; F: mirror mounted on flipper; AOM: acousto-optic modulator; PMT: photomultiplier tube detector; EMCCD: electron-multiplying charge coupled device; other optical elements not marked are ordinary mirrors and convex lenses. Yellow arrow lines: optical fiber; dark green arrows: the direction of the C field (quantization axis); brown arrows: RF pulse (the polarization direction of the magnetic field is plotted); other arrow lines: lasers of different wavelengths (colors used are the same as in Figure \ref{fig:el})
    }
  \label{fig:os}
\end{figure}

The core of the whole system is a linear quadrupole trap (see Figure \ref{fig:os}). Each electrode is divided into three segments:
the middle segment with length of 40 mm that acts as radio frequency (RF) electrodes for radial confinement,
and the two ends of length 20 mm that act as end-cap (EC) electrodes for axial confinement \cite{han2021,miao2022}. The radius of each electrode is $r_e=7.1$ mm and the distance to the main central axis of the trap is $r_0= 6.2$ mm to obtain the most ideal quadrupole field and to reduce RF heating \cite{de1971}.
The ion trap is installed in a vacuum
chamber \cite{han2021}. Using an ion pump and a titanium sublimation pump, the pressure in the vacuum chamber is below $10^{-9}$ Pa, which reduces the frequency shift and decoherence introduced by collision between ions and background
gas molecules. A six-layer magnetic shielding barrel is installed on a liftable stage, which not only ensures
good magnetic shielding, but also facilitates installation and adjustments.

The schematic of the entire experimental system is shown in Figure \ref{fig:os}. 
We used a tunable diode laser operating at a
wavelength of 423 nm and a single mode diode laser operating at a wavelength of 374 nm as the photo-ionization
laser of calcium atoms. The two laser beams are combined by a polarization beam splitter (PBS). To cool and repump ${}^{40}\mathrm{{Ca}}^+$, two laser beams of wavelength 397 nm and 866 nm are combined by a dichromatic mirror. Two
frequency-quadrupled, tunable diode laser systems (TA-FHG Pro, Toptica) are used to ionize the cadmium atoms
(228 nm) and probe the ${}^{113}\mathrm{{Cd}}^{+}$ (214.5 nm). The power of the 228 nm laser beam is about 1 mW, whereas that
of the 214.5 nm laser beam can be adjusted by a half-wave plate and another PBS. All tunable lasers are locked
to a specific frequency measured by a wavelength meter (HighFinesse, WSU 8-2) via an optical fiber. Flipper
optic mounts are used to ensure that the photo-ionizing lasers (wavelengths 374 nm, 423 nm and 228 nm) can
enter the vacuum chamber only during the ionizing process. A Rochon polarizer and a quarter-wave plate are
used to obtain pure right-handed circularly polarized light ($\sigma^+$), ensuring a cycling transition during detection.
To prepare the cadmium ion in the  ${}^{2}\mathrm{{S}}_{1/2}\ F=0$ state (thus completing the clock transition), we use two acousto-optic
modulators (AOM) to shift the probe laser by 800 MHz for pumping. The 15.2 GHz microwave is fed into
the vacuum cavity through a microwave horn. The polarization direction of the microwave magnetic field and
the C field (quantization axis) direction are at a non-orthogonal angle to ensure that both $\sigma $-transitions and $\pi$-transitions can be detected. The fluorescence signal from ${}^{113}\mathrm{{Cd}}^{+}$ is detected by a photomultiplier tube (PMT).
The ion crystal is imaged by an electron-multiplying charge coupled device (EMCCD) to ascertain its shape and
structure.

\begin{figure}[h]
  \centerline{
  \includegraphics[width=\linewidth]{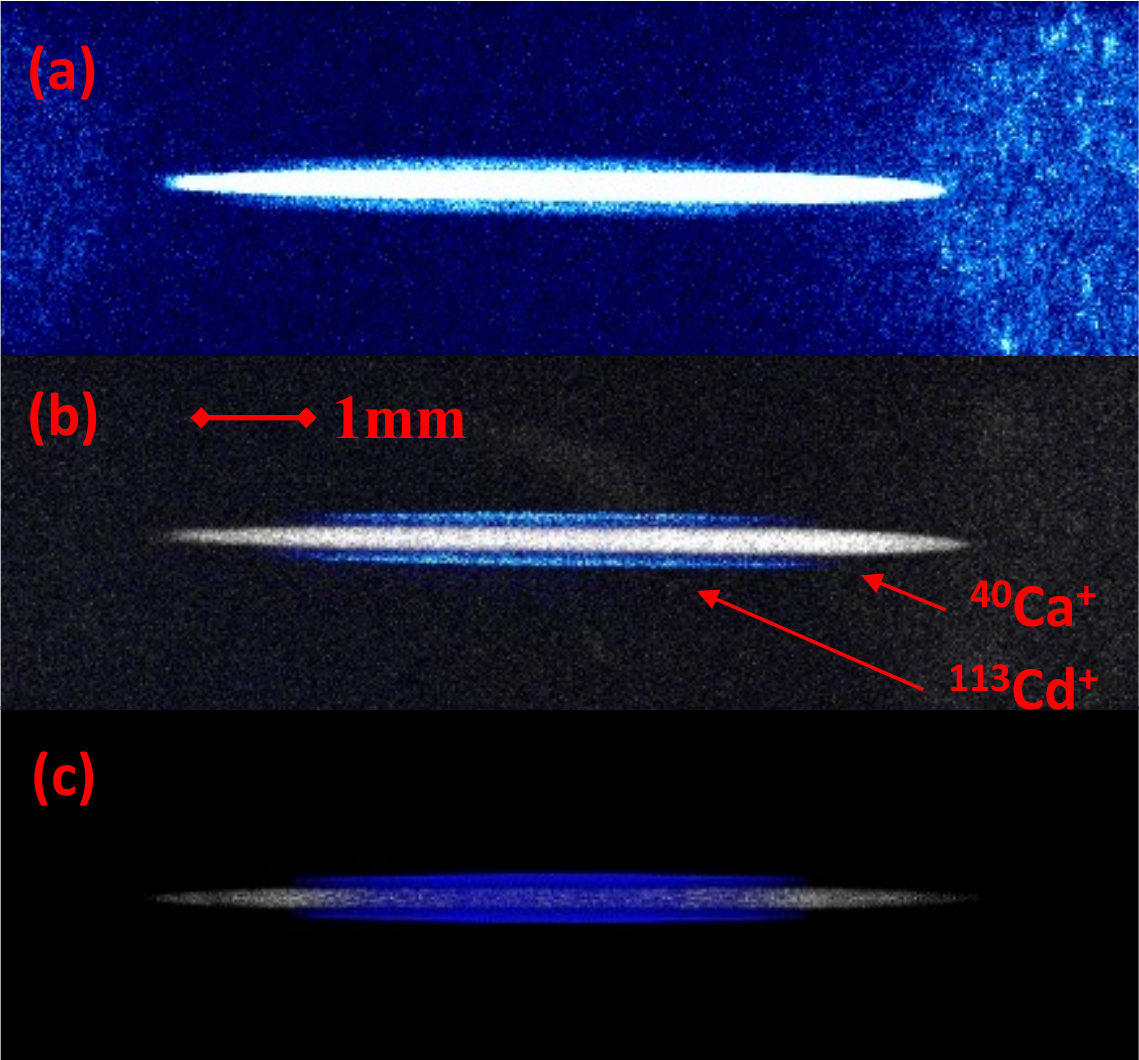}
  }
  \caption{
    Typical image of ${}^{40}\mathrm{{Ca}}^+-{}^{113}\mathrm{{Cd}}^{+}$ bicrystal with $V_{\mathrm{RF}}=170\ \mathrm{V} \ V_{\mathrm{EC}}=8\ \mathrm{V}$. In these pseudocolor images, ${}^{40}\mathrm{{Ca}}^+$ ions appear white and ${}^{113}\mathrm{{Cd}}^{+}$ ions appear blue.
    (a) Image taken by EMCCD without UV filters. 
    Because the power of the 397 nm laser beam is much larger than that for the 214 nm beam, the fluorescence signal of ${}^{40}\mathrm{{Ca}}^+$ is much brighter than that of ${}^{113}\mathrm{{Cd}}^{+}$. Chromatic aberration prevents the two ions from showing a clear image at the same time because the wavelength difference between the two laser beams is nearly double.
    (b) Image synthesized from ion images under different wavelength filters, taking into account the chromatic aberration effects. 
    (c) Molecular dynamics  simulation image of the two-component ion crystal under the same conditions as for (a) and (b).
    }
  \label{fig:ic}
\end{figure}

The ions are confined in a linear quadrupole trap. The frequency of the RF driving voltage is 1.961 MHz, its amplitude
being usually between 150 V and 300 V. The EC voltage for axial confinement is about 5-10 V. To obtain
sympathetically-cooled ion crystals, neutral Ca atoms are first evaporated and ionized and then the ultracold ${}^{40}\mathrm{{Ca}}^+$ crystal is obtained through Doppler cooling. Thereafter, ${}^{113}\mathrm{{Cd}}^{+}$ ions are produced by ionizing evaporated
atoms. By scanning the frequency of the cooling laser (397 nm) and reducing the amplitude of the RF voltage, ${}^{113}\mathrm{{Cd}}^{+}$ ions are sympathetically cooled to the crystal phase \cite{miao2022}. 
A typical image of the two-component ion
crystal (Figure \ref{fig:ic}) displays an ellipsoidal ${}^{40}\mathrm{{Ca}}^+$ crystal located in the center of the trap; the ${}^{113}\mathrm{{Cd}}^{+}$ ions form a thin
shell outside the ${}^{40}\mathrm{{Ca}}^+$ crystal.

Using the zero-temperature charged-liquid model \cite{wine1987ion,h2001st}, the ion density may be estimated using the formula
  \begin{equation}
    \label{eq-d}
    n_i=\frac{\epsilon_0 U_{\mathrm{RF}}^2}{M_i \Omega^2 r_0^4}.
    \end{equation}
  where $\epsilon_0$ denotes vacuum permittivity, $U_{\mathrm{RF}}$  the amplitude of the RF driving voltage, $M_i$  the mass of ion, $\Omega$  the angular frequency RF driving voltage.

  From the EMCCD image, the dimensions of the calcium ion crystal can be measured directly, from which the
  crystal volume can be estimated. However, because the cadmium ion shell is thin, deriving the volume from the
  EMCCD image directly results in a large error. The number of ${}^{113}\mathrm{{Cd}}^{+}$ is calculated by comparing the image
  with that obtained from a molecular dynamics simulation \cite{miao2022}.
  In our sympathetically-cooled ion frequency
  standard experiment, a typical value for the number of ${}^{40}\mathrm{{Ca}}^+$ is 7 500 whereas the number of ${}^{113}\mathrm{{Cd}}^{+}$ is 5 200.

\begin{figure}
      \includegraphics[width=\linewidth]{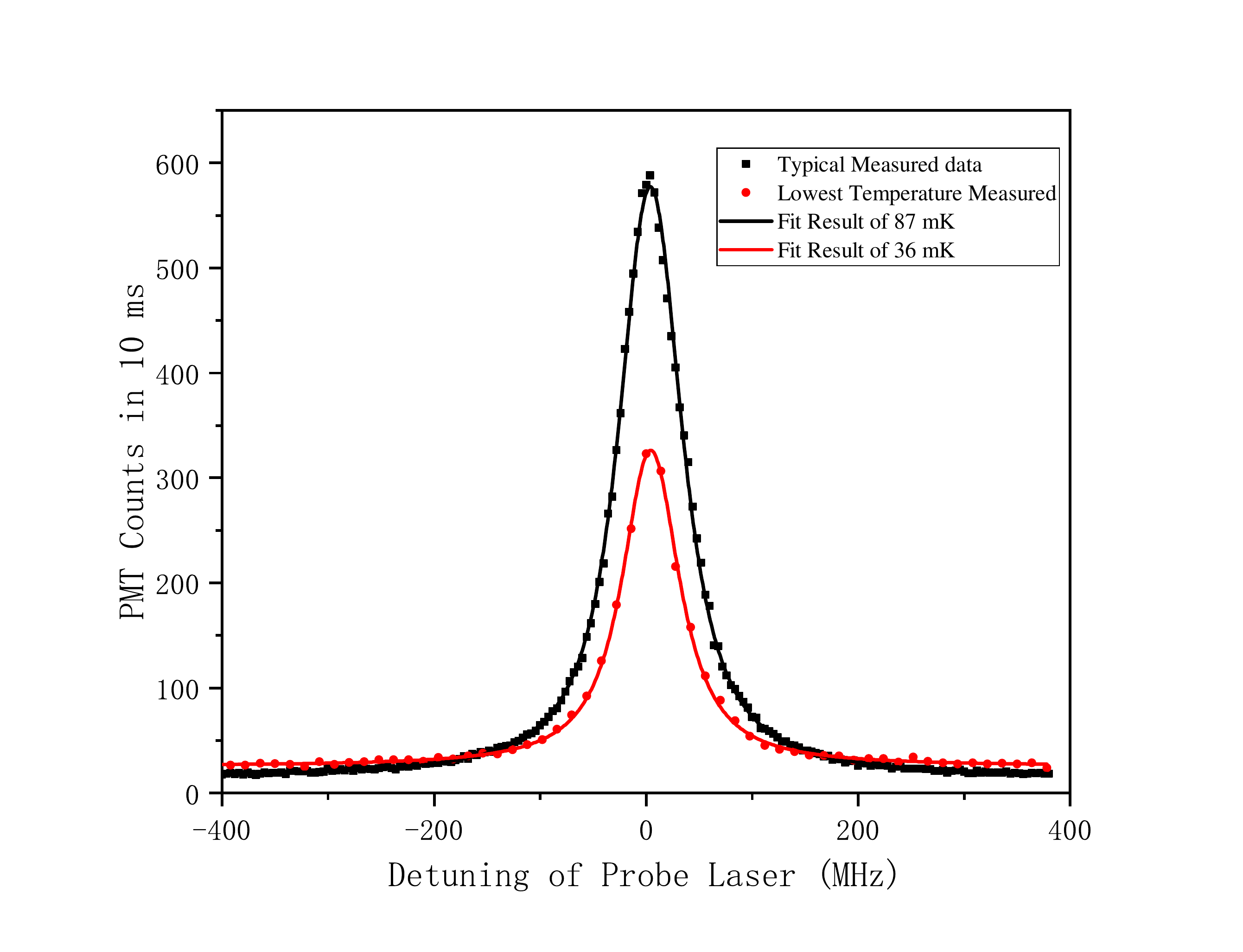}
      \caption{
        Fluorescence spectra of sympathetically-cooled ${}^{113}\mathrm{{Cd}}^{+}$ obtained in experiments exploiting the microwave frequency standard.
        The black line is a fitted curve of data for measurements taken at a typical temperature of 87 mK. Similarly, the red line is a fitted curve of data for the lowest temperature reached of 36 mK.
        }
      \label{fig:te}
    \end{figure}

The temperature of ${}^{113}\mathrm{{Cd}}^{+}$ is an important factor in characterizing the effect of sympathetic cooling. We evaluate
the ${}^{113}\mathrm{{Cd}}^{+}$ temperature by measuring the Doppler broadening of the 214.5 nm ${}^{113}\mathrm{{Cd}}^{+}$ fluorescence spectrum (Figure \ref{fig:te}).
The measured line profiles are fitted with a Voigt curve. The Lorentz linewidth of the Voigt curve
was set at 60.13 MHz, the natural linewidth of the $\mathrm{D}_2$ transition of ${}^{113}\mathrm{{Cd}}^{+}$ \cite{mo2006p}. The temperature can be derived
from the fitted Gaussian linewidth \cite{drewsen2002ion,warrington2002temperature,riehle2006frequency}
    \begin{equation}
      \label{eq-t}
      T=\frac{M_i c^{2}}{8 \ln 2 \cdot k_{\mathrm{B}}}\left(\frac{v_{{G}}}{v}\right)^{2},
      \end{equation}
    where $c$ denotes the speed of light in vacuum, $k_{\mathrm{B}}$  the Boltzmann constant, $v_{{G}}$ the Gaussian linewidth, and $v$  the resonance frequency of the $\mathrm{D}_2$ transition of ${}^{113}\mathrm{{Cd}}^{+}$. In our sympathetically-cooled ion microwave frequency
    standard experiments, a typical Gaussian broadening is 27.6(5) MHz, corresponding to a temperature of 87(3)
    mK. We studied the effects of different RF and EC voltages, ion ratios \cite{miao2022}, detuning of the 397 nm laser beam,
    and other factors on the temperature of cadmium ions. The lowest temperature was 36(5) mK, corresponding to
    Gaussian linewidth of 17.8(1.3) MHz. The temperature is an order of magnitude smaller than that registered in
    our
     previous laser-cooled cadmium frequency standard (654 mK) \cite{Cd2021}.

\section{Clock Signal and its Stability}
\begin{figure}
  \includegraphics[width=\linewidth]{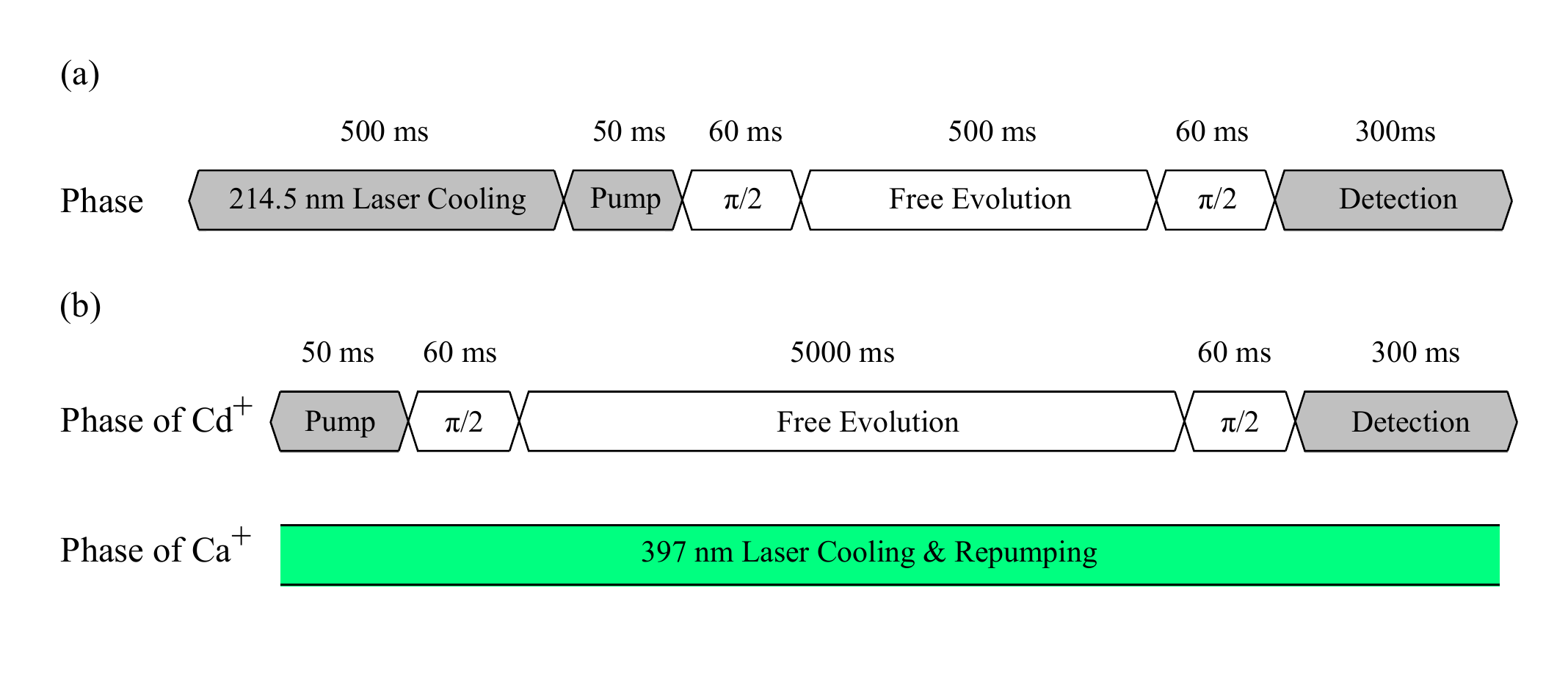}
  \caption{
    Timing sequences for Ramsey's separated oscillatory field method for (a) a traditional laser-cooled  ${}^{113}\mathrm{{Cd}}^{+}$ frequency standard \cite{Cd2021} and (b) a sympathetically-cooled ${}^{113}\mathrm{{Cd}}^{+}$ frequency standard. Dead time marked with a gray background. A 20 ms duration waiting for a circuit response is not shown in the figure.
    }
  \label{fig:ts}
\end{figure}

\begin{figure}
  \includegraphics[width=\linewidth]{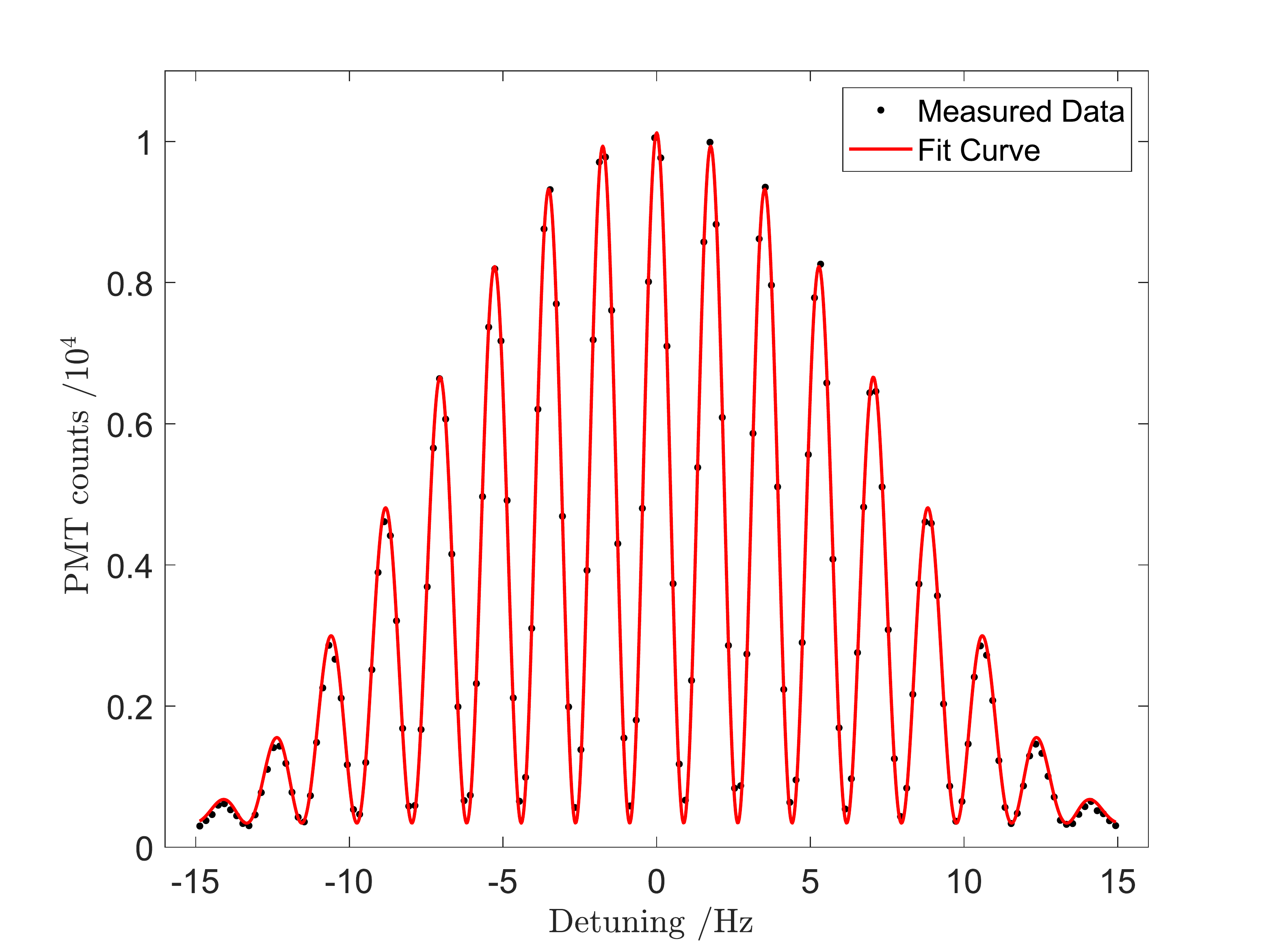}
  \caption{   
Typical high signal-to-noise ratio Ramsey fringe obtained from our sympathetically cooled ion frequency standard operating at a  clock transition of 15.2 GHz with a free evolution time of 500 ms, two phase-coherent microwave pulses of 60 ms, and a fluorescence signal integration time of 300 ms.
    }
  \label{fig:rf}
\end{figure}

After cooling the trapped ions, we obtained the Rabi and Ramsey fringes of the clock transition signal by microwave
interrogation. Timing sequences for the Ramsey spectroscopy in the traditional laser-cooled ${}^{113}\mathrm{{Cd}}^{+}$ frequency standard \cite{Cd2021} and in the sympathetically-cooled ${}^{113}\mathrm{{Cd}}^{+}$ frequency standard are presented in Figure \ref{fig:ts}. 
A typical Ramsey fringe is shown is Figure \ref{fig:rf}. To see the shape of Ramsey fringes clearly, the free evolution time
of
 was chosen to be 500 ms instead of 5 000 ms as noted in Figure \ref{fig:ts}(b).

 During the sympathetic-cooling period of the timing sequence, we measured the loss rate of ions \cite{miao2022}.
 The lifetime of ${}^{40}\mathrm{{Ca}}^+$ was measured from the 397 nm fluorescence decay and was estimated to be more than one day. 
As for ${}^{113}\mathrm{{Cd}}^{+}$, we measured the signal strength at half-height of the Ramsey fringe decay, which falls of exponentially.
The lifetime ($1/e$) was estimated to be up to about $2\times 10^4$ seconds, three times longer than that in a
typical laser-cooling scheme \cite{Cd2021}.

Compared with traditional laser-cooled ion frequency standards, the sympathetically-cooled ion frequency standard
removes laser cooling process from the timing sequence. The dead time per loop (for Ramsey method) is reduced
from 870 ms to 370 ms so that the Dick effect that appears because of the dead time is suppressed.

Moreover, sympathetic cooling can ensure that the cadmium ions are not heated by the RF during microwave
interrogation. The SODFS caused by the thermal motion of ions is effectively suppressed in the sympathetic-cooling
scheme thereby allowing the free evolution time to be extended. The frequency stability of a frequency
standard
 is characterized by the Allan deviation and its theoretical limit is obtained from \cite{riehle2006frequency}
\begin{equation}
  \label{eq-ad}
  \sigma_y(\tau)=\frac{1}{K_0}\frac{1}{Q}\frac{1}{SNR}\sqrt{\frac{T_c}{\tau}},
  \end{equation}
for which $K_0$ is a constant of order unity and $K_0=\pi$ for the Ramsey case; $Q$ is the quality factor $Q=\nu_0/\Delta\nu$, where $\nu_0$ is the clock transition frequency and $\Delta\nu$ is the full width at half maximum of the Ramsey peak; $SNR$ denotes the signal-to-noise ratio, $T_c$  the total cycle time, and $\tau$ the averaging time of stability. In Equation \ref{eq-ad}, $Q$ and $T_c$ are related to the free evolution time $T_f$, specifically, $Q=\nu_0/\Delta\nu=2 T_f \nu_0$ and $T_c= T_f+T_d+2T_p$. In our experiments, dead time is $T_d=370\ \mathrm{ms}$, and the duration of the $\pi/2$ pulse is $T_p=60\ \mathrm{ms}$. When the free evolution 
time 
is much longer than either the dead time or the pulse time, the ultimate frequency stability $\sigma_y$ follows that $\frac{\sqrt{T_c}}{Q}=\frac{1}{2\nu_0}\frac{1}{\sqrt{T_f}}$ dependence given the same $SNR$. Therefore, extending the free evolution time is beneficial in
reducing the theoretical limit of the Allan deviation.

\begin{figure}
  \includegraphics[width=\linewidth]{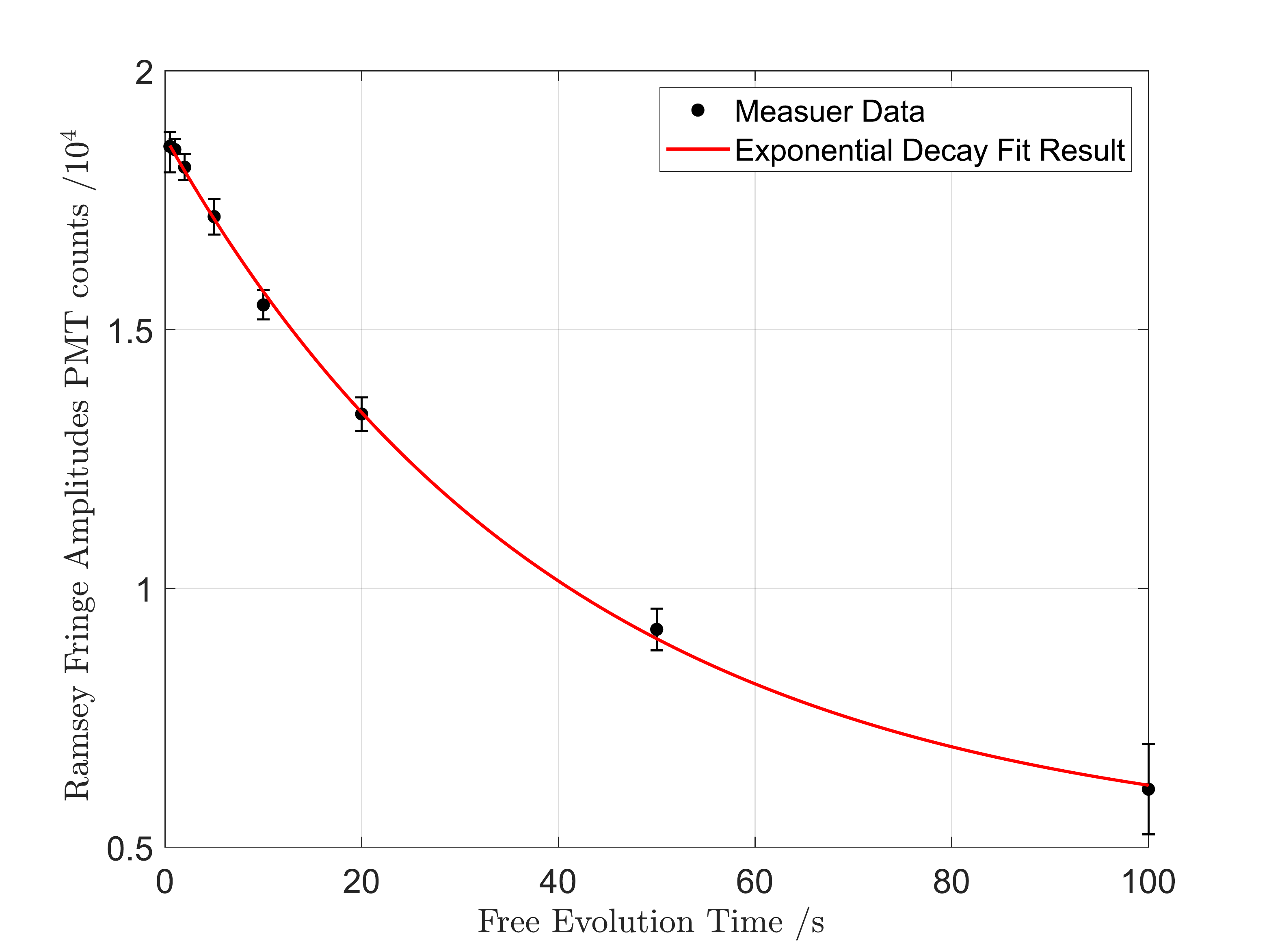}
  \caption{   
    Measurement of Ramsey fringe amplitudes corresponding to different free evolution times. The amplitude is calculated from a cosine fit to the central peak of the Ramsey fringe.
    }
  \label{fig:lt}
\end{figure}

However, as the free evolution time increases, the peak-to-peak value of the Ramsey fringes decreases because
of decoherence in the ${}^{113}\mathrm{{Cd}}^{+}$ crystal. Noise also increases because of magnetic field fluctuations and ion collisions,
among other factors. The reduction of the $SNR$ limits the extension of the free evolution time. To obtain
a suitable free evolution time, we measured the coherence lifetime of sympathetically cooled ${}^{113}\mathrm{{Cd}}^{+}$ (Figure \ref{fig:lt}). The exponential decay fitting gives a $1/e$ lifetime of 41 s. Nonetheless, the measurement time is up to one and a half hours. Considering signal reduction caused by particle loss ($1/e$ lifetime $2\times 10^4$ s), the coherence lifetime should be slightly longer than the 41 s duration obtained by fitting. Taking the above factors into consideration, the free evolution time of closed-loop locking is set at 5 s.

\begin{figure}
  \includegraphics[width=\linewidth]{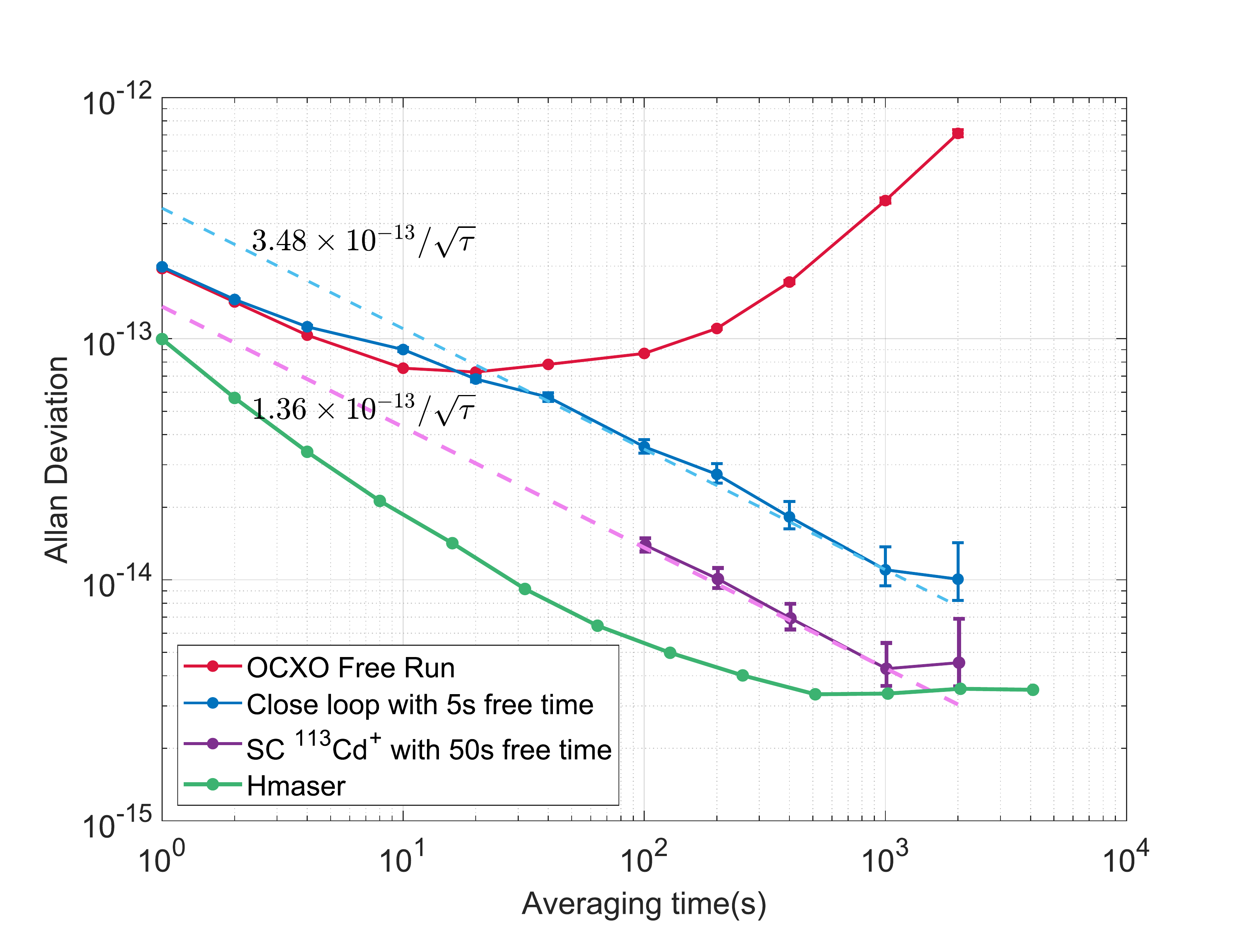}
  \caption{ 
    Allan deviations of the sympathetically-cooled ${}^{113}\mathrm{{Cd}}^{+}$ microwave frequency standard. 
    The solid red line is the stability of the free-running OCXO measured by a phase noise analyzer. 
    The solid blue line is the stability of the OCXO locked on the Ramsey fringe of sympathetically-cooled ${}^{113}\mathrm{{Cd}}^{+}$ with a free evolution time of 5 s, measured by the phase noise analyzer. 
    The purple solid line is the stability derived from the Ramsey fringe error signal obtained from frequency hopping  measurements. 
    The light blue and purple dashed lines are the $\tau^{-1/2}$ fits of the solid lines. 
    The green solid line is the fractional frequency stability of the hydrogen maser (MHM 2010)  used in our laboratory.
    }
  \label{fig:ad}
\end{figure}

The
 local oscillator (LO) of the frequency standard is an oven controlled crystal oscillator (OCXO) with a frequency
stability of $2 \times 10^{-13}$ at 1 s (Figure \ref{fig:ad}, red line).
A microwave synthesizer (8257D, Agilent) converts the LO's 10 MHz signal to 15.2 GHz. By frequency-hopping at the left and right half-heights of the central peak ($\nu_0\pm\Delta\nu/2$), the error signal is obtained by taking the difference between the left and right signal strengths. The
error signal is fed to a digital proportional-integral-differentiation (PID) controller to generate a feedback voltage
to the LO. Once the LO is controlled by the ${}^{113}\mathrm{{Cd}}^{+}$, its output of 10 MHz is fed to a phase noise analyzer
(5120A, Symmetricom) referenced to a hydrogen maser (MHM 2010, Microsemi) for characterization.

The stability measurement of closed-loop locking is marked as a solid blue line in Figure \ref{fig:ad}. The free evolution
time of Ramsey's separated oscillatory fields is 5 s and the feedback loop time is 11 s due to frequency-hopping.
The Allan deviation for times shorter than the loop time is determined using the LO whereas that for longer than the loop time is determined by the ions, indicating a short-term frequency stability of $3.48\times 10^{-13}/\sqrt{\tau}$. The measurement lasted for five hours (18 047 s) and the Allan deviation at 2 000 s is $1.01\times 10^{-14}$. Longer measurement times did not result in better stability, possibly because of a reduced $SNR$ from ion loss.

The short-term frequency stability seems not to have improved much compared with previous results from laser-cooling
scheme ($4.2\times 10^{-13}/\sqrt{\tau}$ in \cite{Cd2021}). The main reason is that the previous results \cite{Cd2013,Cd2015,Cd2021} characterized the stability using a modified Allan deviation, which occurs usually in the presence of fast noise processes \cite{r2008ph,r2008h}, such as those produced by ultrastable lasers rather than atomic clocks. For white frequency noise ($\sigma_y\propto\tau^{-1/2}$), Allan deviation is $\sqrt{2}$ times the modified Allan deviation.

To evaluate the limit of short-term stability, we need to consider the Dick effect and the influence of various types
of noise. We have measured the phase noise power spectral density of our LO at 15.2 GHz via a phase noise
analyzer (53100A, Microchip). Bringing in the experimental parameters of our closed-loop measurements, we
obtain an Allan deviation for the Dick effect contribution of $3.5\times 10^{-14}/\sqrt{\tau}$, one-tenth that for laser-cooling scheme ($3.43\times 10^{-13}/\sqrt{\tau}$ in \cite{Cd2021}). 

For a multi-ion frequency standard, the system noise, including quantum projection noise, shot noise, and pump
noise \cite{itano1993q}, can be calculated from the Ramsey fringes and the number of ions:
\begin{equation}
  \label{eq-n}
  \begin{aligned} 
 &\sigma_{\mathrm{proj}}=\frac{1}{2}K\sqrt{\eta N},\\
 &\sigma_{\mathrm{shot}}=\sqrt{\frac{S+s}{2}},\\
 &\sigma_{\mathrm{pump}}=\frac{\sqrt{N\eta(1-\eta)}K}{2},
  \end{aligned}
  \end{equation}
where $S$ denotes the maximum of Ramsey fringes, $s$  the minimum of Ramsey fringes, $N$  the number of the ${}^{113}\mathrm{{Cd}}^{+}$, $K$  the number of photons stimulated radiated per ion with $K=(S-B)/N$, $B$  the PMT background counts without ions, and $\eta$  the pump efficiency with $\eta=1-(s-B)/(S-B)$.
Summing the squares of each noise gives the total system noise, which contributes $1.38\times 10^{-13}/\sqrt{\tau}$ to the Allan deviation. The rest of the Allan deviation comes from
technical noise, including fluctuations in laser frequency and power, temperature fluctuations, and circuit noise
during the PID process.

To explore the limit of short-term frequency stability, we extend the free evolution time to its extreme of 50 s.
Over this period, the instability of the LO has a substantial influence, and hence we adopted an indirect measurement
method. We used the hydrogen maser (MHM 2010) as a LO, and calculated the relative frequency difference
through the error signal, thereby obtaining the Allan deviation indirectly. The $1.6 \times 10^4$ s  measurement is marked as a
purple dashed line in Figure \ref{fig:ad}. The short-term frequency stability of the measurement is $1.36\times 10^{-13}/\sqrt{\tau}$. For
comparison, the fractional frequency stability of the hydrogen maser is also shown as the green line in Figure \ref{fig:ad}.
There is still room for improvement in our sympathetically-cooled ion frequency standard.

\section{Uncertainty Evaluations}

\begin{figure*}
  \includegraphics[width=\linewidth]{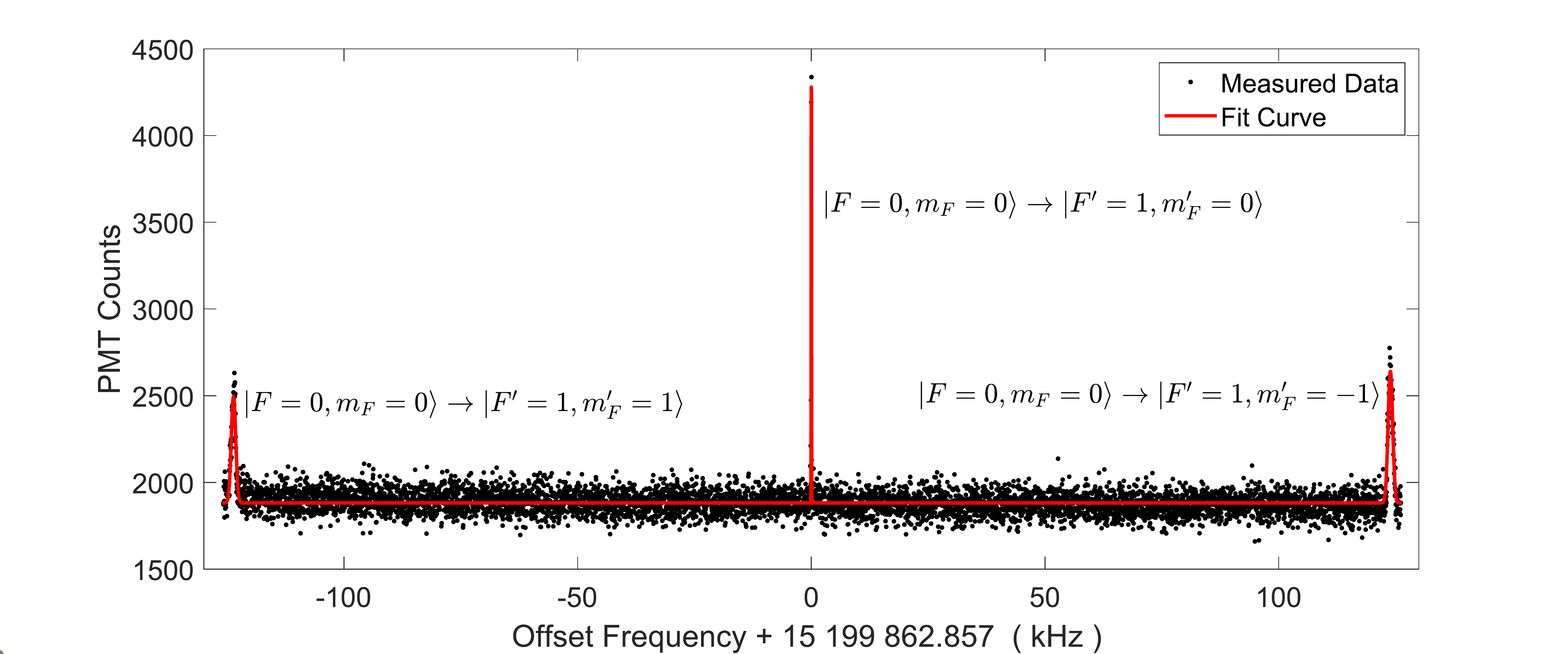}
  \caption{ 
    Rabi fringe of the ground state hyperfine transition ($\ket{F=0, m_F=0}\rightarrow\ket{F'=1, m'_F=0,\pm 1}$). 
    The three peaks from left to right  correspond to the $m_F=0\rightarrow 1$, $m_F=0\rightarrow 0$ and $m_F=0\rightarrow -1$ transition, respectively. The power of the 15.2 GHz
  $\pi$  pulse is $-9$ dBm;
   a range of 25.2 kHz was scanned in steps of 40 Hz.
    }
  \label{fig:cm}
\end{figure*}

We first discuss the second-order Zeeman frequency shift (SOZFS), which is the main source of systematic uncertainty
in this work. We measured the Rabi fringe of the magnetic-field-sensitive transition ($\ket{F=0, m_F=0}\rightarrow\ket{F'=1, m'_F=\pm 1}$, Figure \ref{fig:cm}), and the SOZFS, the fractional values of which are given by
\begin{equation}
  \label{eq-soz}
  \begin{aligned} 
  \frac{\delta \nu_\mathrm{SOZFS}}{\nu_0}&=\frac{\left(g_J-g_I\right)^2\mu_B^2}{2 h^2 \nu_0^2}B^2\\
&=\frac{1}{2\nu_0^2}\left(\frac{g_J-g_I}{g_J+g_I}\right)^2\times \left(\nu_{0-1}-\nu_{0+1}\right)^2,
  \end{aligned} 
  \end{equation}
where $\nu_0$ denotes the ground-state hyperfine splitting of ${}^{113}\mathrm{{Cd}}^{+}$, $\mu_B$  Bohr magneton, $h$  Planck constant, $g_J$ and $g_I$ denote the respective Land\'e $g$-factors of the electron and the nucleus, having values $g_J=2.002291(4)$ \cite{yu2020gj} and $g_I=0.6223009(9)\times 10^{-3}$ \cite{s1972gI}, and $\nu_{0-1}$ and $\nu_{0+1}$ denote the frequencies of magnetic-field-sensitive transition
we measured. The frequency difference between the two peaks of the magnetic-field-sensitive transition is
247 488(10) Hz, and hence the magnetic field strength is 8828.3(4) nT with a fractional SOZFS of  $1.32\ 391(11)\times 10^{-10}$. The calculated result is larger that presented in \cite{han2021} because the C field in the experiment is larger than
the ideal case.

Before calculating the SODFS, we should analyze in advance the contributions to the ion velocity of the different
motions. The motion of ions in a Paul trap can be decomposed into secular motion, micromotion induced by
the RF field, and the excess micromotion induced by the phase difference between the two RF electrodes or additional
electrostatic fields \cite{b1998mi}.
For the first two, their contributions to SODFS can be calculated using \cite{p1999h}
\begin{equation}
  \label{eq-sod1} 
  \frac{\delta \nu_\mathrm{SODFS-sm}}{\nu_0}=-\frac{3 k_B T}{2 M c^2}\left(1+\frac{2}{3}N_d^K\right),
  \end{equation}
  where $T$ denotes the temperature of the ion ensemble (here, we can use 87(3) mK obtained in Section 2), and $N_d^K=3$ set the SODFS coefficient associated with the micromotion. The fractional SODFS is calculated to be $-3.21(11)\times 10^{-16}$. 
  However, for sympathetically cooled large-size ion crystals in a linear Paul trap, excess micromotion
  provides the dominant contribution to the SODFS. We can suppose the interaction of calcium ions
  with cadmium ions as a constant electrostatic field, forcing the cadmium ions to deviate from the central axis of
  the Paul trap. Its contribution to the SODFS may be calculated from
\begin{equation}
  \label{eq-sod2} 
  \frac{\delta \nu_\mathrm{SODFS-ex}}{\nu_0}=-\frac{1}{16}\frac{q^2 \Omega^2 u^2}{c^2},
  \end{equation}
where $q$ denotes a dimensionless quantity describing the motion of the ion with $q=2 Q_i U_{\mathrm{RF}}/(M_i\Omega^2r_0^2)$, $Q_i$ and $M_i$ denote the charge and mass of ${}^{113}\mathrm{{Cd}}^{+}$, and $u$ denotes the distance of the ion from the central axis of the Paul
trap. This distance can be estimated from the ion crystal size using the EMCCD images; hence, the fractional
SODFS contributed by excess micromotion is calculated to be $-2.4(3)\times 10^{-14}$. The SODFS contributed by the excess micromotion also limits the number of trapped ions. We cannot blindly increase the number of ${}^{113}\mathrm{{Cd}}^{+}$ to
improve the $SNR$, because that will increase the SODFS at the same time.

We discuss the Stark frequency shift separately depending on the source. Because all of the 214.5 nm laser beams
are blocked by mechanical shutters during the microwave interrogation, its contribution to light frequency shifts
is negligible. However, because both cooling and repump lasers remain on during the microwave interrogation,
the sympathetic-cooling scheme will introduce additional light frequency shifts compared with those from laser
cooling; see \cite{han2021} for a detailed discussion. Substituting in the light intensity for our experiment (which is the
same as that described in \cite{han2021}), the number of ${}^{40}\mathrm{{Ca}}^{+}$ (7 500), and the surface area of the calcium ion crystals (6.9 mm${}^2$), we obtain the light frequency shifts contributed by the laser beams, the fluorescence being $5.4(5)\times 10^{-17}$ and $1.2(1.2)\times 10^{-21}$,respectively.

The Stark shifts generated by the electric field of the ion trap, may be estimated from the average of the square
of the ion velocities $\left<V^2\right>$ \cite{b1998mi}, which may be calculated from the SODFS, specifically $\delta \nu_{\mathrm{SODFS}}/\nu_0=-\left<V^2\right>/(2c^2)$, and given by
\begin{equation}
  \label{eq-s} 
  \frac{\delta \nu_\mathrm{S-trap}}{\nu_0}=\frac{\sigma_S}{\nu_0}\left( \frac{m \Omega}{Q_i^2}\right)^2\left<V^2\right>,
  \end{equation}
where $\sigma_S$ denotes the static Stark shift and is calculated to be $3.99(11)\times 10^{-12}$Hz$/($V$/$m$)^2$ \cite{yu2017stark}. Therefore, the
Stark frequency shifts contributed by the ac quadrupole electric field (corresponding to the secular motion and
micromotion) and the electrostatic field generated by the calcium ions (corresponding to the excess micromotion)
are $3.14(14)\times 10^{-18}$ and $2.3(3)\times 10^{-16}$, respectively.

The Stark frequency shift from the blackbody radiation contribution (BBRS) can be obtained from \cite{yu2017stark}. Considering
the uncertainty of the calculation results and the fluctuation in room temperature of  $300\pm 10$ K, the fractional
BBRS shift and its uncertainty is $-1.8(2)\times 10^{-16}$. Similarly, the Zeeman frequency shift of the blackbody
radiation (BBRZ) along with its uncertainty is calculated to be $-9.8(7)\times 10^{-18}$ \cite{han2019BBRZ}.

For the ac Zeeman shift contributed by the quadrupole field, the ac magnetic field vanishes if the trap structure
is physically and electrically symmetric \cite{o2006single}. In estimating its uncertainty, we can give an upper bound to the ac
magnetic field of $10^{-10}$ T obtained from the electric field average \cite{b1998mi} and the RF frequency, which contributes
an ac Zeeman shift less than  $2\times 10^{-20}$.

The gravitational redshift has a magnitude of $1.1\times 10^{-16}$ 
per meter of the change in height at sea level 
\cite{chou2010optical}.
The
altitude of the experimental system in our laboratory is $43\pm 1$ m \cite{Cd2021} so that the fractional frequency shift resulting
from the gravitational redshift is estimated to be $4.73(11)\times 10^{-15}$ compared to sea level.

For laser-cooled ions, the pressure frequency shift is negligible for an ultrahigh vacuum (and the same situation
for sympathetic cooling cases). Assuming the pressure shift coefficient for ${}^{113}\mathrm{{Cd}}^{+}$  is of the same order as for  ${}^{171}\mathrm{{Yb}}^{+}$ \cite{park2007yb+} and  ${}^{199}\mathrm{{Hg}}^{+}$ \cite{chung2004buffer}, and considering the pressure in the vacuum chamber is below $1\times 10^{-9}$ Pa and
mostly contributed by He and H${}_2$, the uncertainty in the pressure frequency shift is no more than $10^{-17}$.

\begin{table*}
  \caption{Estimation of Systematic Frequency Shifts and
  Uncertainties}
  \label{ta:un}
  \begin{ruledtabular}
  \begin{tabular}[t]{l d l d l}
    \multicolumn{1}{l}{Shift} & \multicolumn{2}{l}{Fractional Magnitude of Effect} & \multicolumn{2}{l}{Fractional Uncertainties}  \\ 
    \hline
    SOZFS
         &1.32\ 391   & $\times 10^{-10}$
    &   1.1         & $\times 10^{-14}$
    \\
    SODFS by secular and micro- motion 
      & -3.21      & $\times 10^{-16}$ 
    &    1.1         & $\times 10^{-17}$
    \\
    SODFS by excess micromotion 
      &  -2.4        & $\times 10^{-14}$ 
    &    3           & $\times 10^{-15}$
    \\
    light shift by 214.5 nm 
       &  0         & 
    &     0           & 
    \\      
    light shift by 397 \& 866 nm (laser)  
       &  5.4        & $\times 10^{-17}$ 
    &    5           & $\times 10^{-18}$
    \\
    light shift by 397 \& 866 nm (flou.)
        &  1.2        & $\times 10^{-21}$ 
    &    1.2         & $\times 10^{-21}$
    \\
    ac Stark by quadrupole field 
       &  3.14       & $\times 10^{-18}$ 
    &     1.4         & $\times 10^{-19}$
    \\
    dc Stark by Ca${}^{+}$ electrostatic field 
       &  2.3        & $\times 10^{-16}$ 
    &    3          & $\times 10^{-17}$
    \\
    BBRS
    &  -1.8        & $\times 10^{-16}$ 
    &     2           & $\times 10^{-17}$
    \\
    BBRZ
     &  -9.8        & $\times 10^{-18}$ 
    &     7           & $\times 10^{-19}$
    \\
    ac Zeeman by quadrupole field
     & 0             & 
    & < 2           & $\times 10^{-20}$
    \\
    gravitational red shift   
      & 4.73        & $\times 10^{-15}$
    &     1.1         & $\times 10^{-16}$
    \\
    pressure shift 
       &  0      & 
    & < 1           & $\times 10^{-17}$
    \\
    Total 
     &  1.32\ 371  & $\times 10^{-10}$ 
    &     1.1         & $\times 10^{-14}$
    \\
\end{tabular}
\end{ruledtabular}
\end{table*}

From the systematic frequency shifts and corresponding uncertainties listed in Table \ref{ta:un}, the SOZFS contributes
most of the uncertainty and is larger than previous results \cite{Cd2021}, probably because the magnetic shield barrel deteriorates
from long-term use or the effect of geomagnetic fluctuations during the experiment. The good news
is that the SODFS is half smaller that of the laser cooling scheme, and the additional light frequency shift introduced
by sympathetic cooling can be ignored. Taking into account the type A uncertainty ($\sigma_y=1.01\times 10^{-14}$ @ 2000 s), the total uncertainty for the sympathetically-cooled ${}^{113}\mathrm{{Cd}}^{+}$ frequency standard is $1.5\times 10^{-14}$, which is
superior to the level obtained from laser-cooled cadmium ions frequency standard ($1.8\times 10^{-14}$ in \cite{Cd2021}).

\section{Conclusions and Discussions}
We built the first Paul-trap-based sympathetically-cooled ion microwave frequency standard, based on ${}^{113}\mathrm{{Cd}}^{+}$
ions cooled by coolant ions of laser-cooled ${}^{40}\mathrm{{Ca}}^{+}$.
 Sympathetically-cooled cadmium ions can attain ultra-low
temperatures of 36 mK with coherent lifetimes exceeding 40 s. In adopting the sympathetic-cooling method, the
dead time can be compressed from 870 ms to 370 ms, and the free evolution time of the closed loop can be extended
from 0.5 s to 5 s so that the Dick effect is suppressed to one-tenth that of laser-cooling scheme (from $3.43 \times 10^{-13}/\sqrt{\tau}$ in \cite{Cd2021} to $3.5 \times 10^{-14}/\sqrt{\tau}$ in this study). We achieved a closed-loop locking experiment with a short-term frequency stability of $3.48 \times 10^{-13}/\sqrt{\tau}$, close to the level of the laser-cooled mercury ions microwave
clock ($3.3 \times 10^{-13}/\sqrt{\tau}$ in \cite{Hg1998}), although the ${}^{113}\mathrm{{Cd}}^{+}$ clock's transition frequency (15.2 GHz) is about $1/3$ that of ${}^{199}\mathrm{{Hg}}^{+}$ (40.5 GHz).  
The fractional uncertainty is $1.5\times 10^{-14}$, better than that for laser-cooled cadmium ions frequency
standard ($1.8\times 10^{-14}$ in \cite{Cd2021}). The improvement in accuracy is mainly because of the reductions in type
A uncertainties and SODFS uncertainties. In exploring the limits of sympathetically-cooled ion microwave frequency
standard, the stability of the 50 s-free-evolution was measured indirectly to be $1.36 \times 10^{-13}/\sqrt{\tau}$ with a
hydrogen maser as LO.

In the future, we will focus on the long-term frequency stability of the sympathetically-cooled ion microwave
frequency standard, as it is an important part of the performance of microwave frequency standards and reducing
the type A uncertainty can give higher frequency accuracy. Moreover, ${}^{174}\mathrm{{Yb}}^{+}$ is seen as a replacement for ${}^{40}\mathrm{{Ca}}^{+}$ in experiments to attain higher performance \cite{miao2021}. Miniaturization is also another consideration because
the sympathetically cooled ion microwave frequency standard holds promise in establishing a ground-based time
frequency reference for future satellite navigation systems.

\begin{acknowledgments}
  The authors thank Dr. Zheng-Bo Wang, Dr. Kai Miao, Dr. Chen-Fei Wu, Dr. Ya-ni Zuo,  Li-Ming Guo, Hua-Xing Hu, Wen-Xin Shi, Tian-Gang Zhao and Ying Zheng for helpful assistance and discussions. 
  This work is supported by National Key Research and Development Program of China (2016YFA0302100,
  2021YFA1402101), and National Natural Science 
  Foundation of China (12073015).
  \end{acknowledgments}

\bibliography{apssamp}

\end{document}